\documentclass[11pt, preprint]{aastex}
\usepackage{subfigure}
\let\tablehead\undefined
\let\tabletail\undefined
\usepackage{supertabular}
\let\affil\undefined
\usepackage{authblk}
\def\ha{H$\alpha~$}
\def\hap{H$\alpha$}
\def\iha{$I_{\textrm{H}\alpha}~$}
\def\ihap{$I_{\textrm{H}\alpha}$}
\def\lha{$[L_{\textrm{H}\alpha}/L_{\textrm{bol}}]~$}
\def\lhap{$[L_{\textrm{H}\alpha}/L_{\textrm{bol}}]$}
\def\id{$I_{\textrm{D}}~$}
\def\idp{$I_{\textrm{D}}$}

\def\rpdp{$R'_{\textrm{D}}$}

\title{\bf Secretly Eccentric: The Giant Planet and Activity Cycle of GJ 328}
\author[1]{Paul Robertson}
\author[1]{Michael Endl}
\author[1]{William D. Cochran}
\author[1]{Phillip J. MacQueen}
\author[2]{Alan P. Boss}
\affil[1]{Department of Astronomy and McDonald Observatory, University of Texas at Austin, Austin, TX 78712, USA; paul@astro.as.utexas.edu}
\affil[2]{Department of Terrestrial Magnetism, Carnegie Institution, 5241 Broad Branch Road, NW, Washington, DC 20015-1305, USA}

\begin{abstract}

We announce the discovery of a $\sim2$ Jupiter-mass planet in an eccentric 11-year orbit around the K7/M0 dwarf GJ 328.  Our result is based on 10 years' worth of radial velocity (RV) data from the Hobby-Eberly and Harlan J. Smith telescopes at McDonald Observatory, and from the Keck Telescope at Mauna Kea.  Our analysis of GJ 328's magnetic activity via the Na I D features reveals a long-period stellar activity cycle, which creates an additional signal in the star's RV curve with amplitude 6-10 m/s.  After correcting for this stellar RV contribution, we see that the orbit of the planet is more eccentric than suggested by the raw RV data.  GJ 328b is currently the most massive, longest-period planet discovered around a low-mass dwarf.

\end{abstract}

\begin{document}

\section{\bf Introduction}

The discovery of planets around nearby stars has become essentially routine, and exoplanets are now understood to be common around nearly every kind of star.  Low-mass planets, in particular, appear to be virtually ubiquitous even for low-mass \citep{howard12} and low-metallicity \citep{buchhave12} stars, as shown in recent \emph{Kepler} results.  However, the one population of planets which remains sparse despite the recent explosion of exoplanet discoveries is gas giant planets in long-period orbits around low-mass stars.

The under-abundance of giant planets around low-mass stars is understood within the core accretion model as the result of less massive protoplanetary disks and longer dynamical timescales for such stars.  As a result, protoplanetary cores fail to acquire enough material to undergo runaway accretion within the disk lifetime \citep[e.g.][]{laughlin04,alibert11}.  Giant planets do exist in both close \citep[e.g. GJ 876b,c][]{marcy98,delfosse98,marcy01} and distant (GJ 676Ab \citep{forveille11}, GJ 832b \citep{bailey09}) orbits around M stars.  Generally, though, gas giants ($M \sin i > 0.3 M_J$) are increasingly rare around lower-mass stars, particularly below $M_* = 0.85 M_{\odot}$ \citep{johnson10,wright11}.  In particular, close-in gas giants are exceedingly rare around M stars \citep{endl06,muirhead12,swift13}, in agreement with the predictions of core accretion simulations.  It is important to note, though, that to date there is not yet sufficient observational data to confirm predictions that low-mass stars are deficient in giant planets at \emph{all} orbital radii.  Since it is conceivable that planet formation through gravitational instability may contribute to the long-period Jovian population for M stars \citep{boss98,boss06,boss11}, it is essential to probe the space beyond $\sim 1$ AU around M dwarfs for exoplanets.

The early indications are that the early M dwarfs have detectable gas giant planets ($M \sin i$ greater than 0.8 Jupiter masses) on relatively short period orbits (semi-major axes less than 2.5 AU) with a frequency of about 5\% \citep{johnson10}. However, gravitational microlensing searches have uncovered giant planets orbiting at greater distances at a rate consistent with a much higher frequency of giant planet companions to early M dwarfs, about 35\% \citep{gould10}. Microlensing has also detected a huge population of Jupiter-mass planets (about two per main sequence star) that are either unbound, or at distances of 10 AU or more from their host stars \citep{sumi11}. \citet{quanz12} showed that given the weak limits imposed by direct imaging surveys, most of these objects are likely to be bound to a star rather than to be free-floating. They suggest that Jupiter-mass planets are likely to be found in increasing numbers out to about 30 AU around M dwarfs. Evidently there is much remaining to be learned about the census of extrasolar giant planets.

The McDonald Observatory M dwarf exoplanet survey \citep[see][for details, including selection criteria]{endl03} is a long-term radial velocity (RV) survey of 100 M stars using the Hobby-Eberly Telescope.  The survey has just finished its 12th year of operation, and has amassed approximately 3,000 high-precision RV measurements over that interval.  We have previously established the paucity of hot Jupiters around M dwarfs \citep{endl06}, and now use the survey primarily to discover low-mass planets and long-period gas giants.  Additionally, we are in the process of analyzing the long-term magnetic activity of our stellar targets \citep{robertson13}.

In this paper, we announce GJ 328b, a Jovian planet around a K/M dwarf at $a \sim 4.5$ AU.  Although the star exhibits a solar-type activity cycle that shifts our measured RVs, we show that the signal of planet b is not caused by this behavior.  Instead, when correcting for stellar activity, we see that the planet follows a more eccentric orbit than indicated by our model to the uncorrected velocities.

\section{\bf Stellar Properties of GJ 328}

In Table \ref{stellar}, we list measured and derived stellar parameters for GJ 328.  Because the absence of an optical continuum makes estimating these properties for low-mass stars difficult with spectral synthesis, we instead rely on photometric calibrations for several items in the table.  We derive a mass of $M_* = 0.69 \pm 0.01 M_{\odot}$ using the \citet{delfosse00} K-band mass-luminosity curve, and a metallicity [M/H] = $0.00 \pm 0.15$ from the \citet{schlaufman10} color-magnitude relation.

With a mass of 0.69 solar masses and $T_{eff} = 3900 \pm 100$ K, GJ 328 lies at the boundary between spectral types K and M.  Various catalogs classify the star between K7 \citep{anderson12} and M1 \citep{kharchenko09}.  For our purposes, it suffices to say that GJ 328 is a ``cool dwarf."  We note that, using the Galactic $UVW$ velocities of \citet{bobylev06}, the probability distributions of \citet{reddy06} indicate GJ 328 is a thin-disk star with greater than 99 percent confidence.

\section{\bf Data}

Our primary data consist of 58 high-resolution spectra of GJ 328 taken with the High Resolution Spectrograph \citep[HRS,][]{tull98} on the 9.2m Hobby-Eberly Telescope \citep[HET,][]{ramsey98} at McDonald Observatory.  The observations span the period between January 2003 and April 2013, for a total observational time baseline of 10 years.  Our observations were taken using a $2\arcsec.0$ fiber imaged through a $0\arcsec.5$ slit, for a resolving power $R = 60,000$.  The queue-schedule system of HET \citep[see][for details of our HET strategy]{cochran04} allows us to observe GJ 328 several times per season, resulting in quasi-unifom phase coverage for long-period planets.  We perform routine wavelength calibration, bias subtraction, flat-fielding, and cosmic ray removal using standard IRAF\footnote{IRAF is distributed by the National Optical Astronomy Observatories, which are operated by the Association of Universities for Research in Astronomy, Inc., under cooperative agreement with the National Science Foundation.} scripts.

In order to obtain high-precision velocities from our spectra, we place an absorption cell filled with molecular iodine (I$_2$) vapor in front of the slit.  The cell is kept at constant temperature and pressure, which results in the superposition of thousands of stable absorption lines over our spectra from 5000-6400 \AA.  We have a high-S/N spectrum taken without the I$_2$ cell, which we deconvolve from the instrumental profile (IP) using the Maximum Entropy Method.  This ``template" spectrum serves as a baseline against which we model the time-variant IP and RV for each star + I$_2$ template.  The deconvolution and modeling for our spectra are handled by our RV extraction code AUSTRAL, the details of which are described in \citet{endl00}.

Our HET/HRS spectra are supplemented with 14 spectra from the Robert J. Tull Coud\'e spectrograph \citep{tull95} on McDonald's 2.7m Harlan J. Smith Telescope (HJST).  We have also acquired 4 spectra using the HIRES \citep{vogt94} spectrograph on the 10m Keck I telescope at Mauna Kea.  These spectra were taken during our Keck CoRoT RV follow-up observations \citep{santerne11} when the CoRoT field was unobservable.  The HJST and Keck observations were also taken through I$_2$ absorption cells, at resolving powers 60,000 and 50,000, respectively.  Again, RV extractions are performed with AUSTRAL.

We present all of our RV data in Table \ref{rv}, and present them as a time series in Figure \ref{rvfit}.  The velocities have been corrected to remove the velocity of the observatory around the solar system barycenter at the time of observation.

\section{\bf RV Analysis}
\label{analysis}

\begin{figure}
\begin{center}
\includegraphics[scale=0.5]{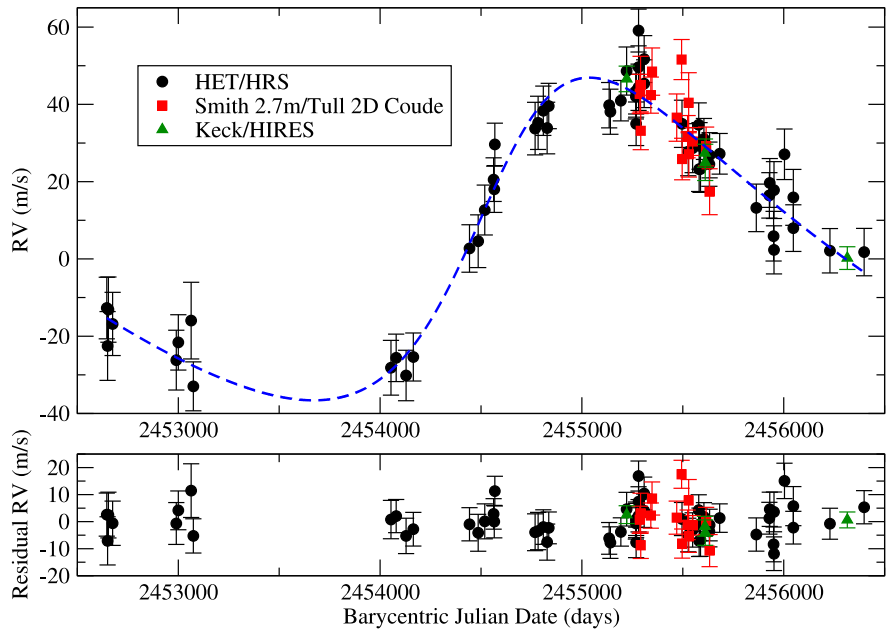}
\caption{We present 10 years of radial velocity data for GJ 328.  The RVs show a clear periodic signal, which we associate with a giant planet.  Our Keplerian model to the data is shown as a dashed blue line.}
\label{rvfit}
\end{center}
\end{figure}

As seen in Figure \ref{rvfit}, our velocities for GJ 328 show large deviations of an apparently periodic nature.  The combined data set display an RMS scatter of 26.7 m/s with an average error of 6 m/s.

\begin{figure}
\begin{center}
\includegraphics[scale=0.5]{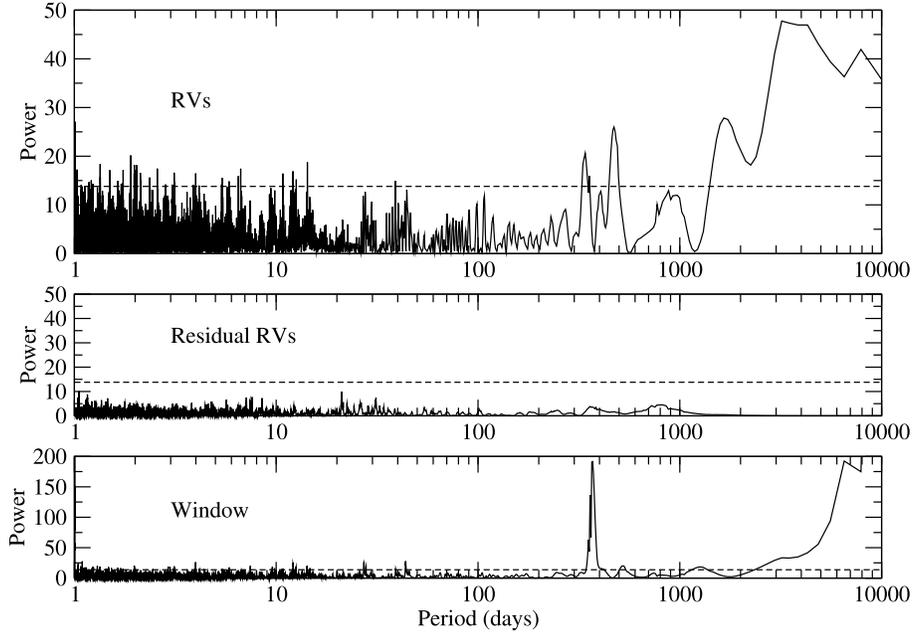}
\caption{Fully generalized Lomb-Scargle periodograms for our RV data.  The top panel shows the power spectrum for our combined RV data set with the strongest peak at $P \simeq 3500$d, while the middle panel gives the periodogram of the residuals to a one-planet fit, and the bottom shows the periodogram of our sampling pattern (the window function).  The dashed horizontal lines represent the power level corresponding to a false alarm probability (FAP) of 0.01, as calculated from Equation 24 of \citet{zk09}.  We note that these FAP levels represent a preliminary estimate, and our formal FAP values are obtained through a bootstrap analysis, which we describe in Section \ref{analysis}.}
\label{rv_ps}
\end{center}
\end{figure}

We began our analysis by computing the fully generalized Lomb-Scargle periodogram \citep{zk09} for our data.  The resulting power spectrum, shown in Figure \ref{rv_ps}, displays a broad, highly significant peak centered at $P = 3700$ days.  We have also computed the window function, the periodogram of our time sampling.  We note that our time sampling is not dense enough to avoid aliasing at very short ($P \sim 1$ day) periods, so our periodograms should be used with caution in the high-frequency regime.

In order to establish the reliability of our periodogram analysis, we compute a false alarm probability (FAP) using three different methods.  The first method assumes the noise distribution in the data is Gaussian, and estimates the probability that a peak of power $P_{\omega}$ would arise at random from a sample of $N$ points.  The formula for this FAP is given in Equation 24 of \citet{zk09}, and we estimate $M$, the number of significant frequencies sampled as $\frac{\Delta f}{\partial f}$, where $\Delta f$ is the range of frequencies in the periodogram and $\partial f$ is the resolution of our computation.  A second estimate, from \citet{sturrock10}, derives the FAP from the excess of power above an expected distribution of power values from Bayesian statistics.  We consider both of these calculations to be preliminary estimates, and note that they agree with each other in all cases discussed herein.  For the 3700-day peak in the RV power spectrum, we compute a preliminary FAP lower than our computational precision, approximately  $1.0 \times 10^{-14}$.

To assign a formal FAP to the signal in Figure \ref{rv_ps}, we conducted a Monte Carlo bootstrap calculation as described in \citet{kurster97}.  We retained the time stamps of the original data set, and assigned at random (with replacement) an RV value from the set to each time.  We computed the periodogram of 10,000 such ``fake" data sets, and took the FAP to be the percentage of those periodograms with a peak having greater power than the original power spectrum at any period.  For the 3700-day signal, our bootstrap calculation produced no false positives in 10,000 trials, giving an upper limit to the FAP of $10^{-4}$.

We begin with the assumption that the variability in the RVs is caused by a planetary companion.  We model the orbit of the planet using both the GaussFit \citep{jefferys88} and SYSTEMIC \citep{meschiari09} software suites.  The observed velocities are fit nicely with the Keplerian orbit of a giant planet.  Our best-fit solution, the parameters of which we list in Table \ref{orbit}, has a period $P = 4100$ days, which corresponds well with the peak in our periodogram when considering the breadth of the peak and the fact that our solution has a significant eccentricity $e = 0.29 \pm 0.04$.  The signal, with an RV amplitude of 42 m/s, corresponds to a planet of minimum mass $M \sin i = 2.5 M_J$ at $a = 4.43$ AU.  The planet, GJ 328b, is therefore a ``cold Jupiter," a gas giant planet which remained at large orbital separation rather than migrating inward.  We show our fit to the data in Figure \ref{rvfit}, as well as the residual RVs after subtracting the model.  Our model results in a reduced $\chi^2$ of 1.12, with a residual RMS scatter of 6.0 m/s.  We note that we see no additional signals in the periodogram of the residual RVs (middle panel, Figure \ref{rv_ps}).

\section{\bf Stellar Magnetic Activity}
\label{activity}

We are currently performing an in-depth analysis of stellar magnetic activity in the stars from our M dwarf RV survey \citep[see][for full details]{robertson13}.  Because stellar activity can cause apparent RV shifts which may obscure \citep{dumusque12} or imitate \citep{queloz01} a Keplerian planet signal, it is essential to consider activity indicators when analyzing RV data.  Although it is not typical for the stellar magnetic behavior of old, quiet stars to create RV signals with magnitudes as large as seen for GJ 328, we have shown in the case of GJ 1170 that stellar activity can occasionally mimic the velocity signatures of giant planets \citep{robertson13}.  We have therefore examined two spectroscopic tracers of stellar activity for GJ 328, and corrected the RV curve for its periodic stellar component.

\subsection{Stellar Activity Analysis}

Because HRS does not acquire the Ca II H \& K lines, we have examined the magnetic activity of GJ 328 using the \ha and Na I D ($\lambda_{\textrm{D1}} = 5895.92$ \AA, $\lambda_{\textrm{D2}} = 5889.95$ \AA) absorption lines from each RV spectrum.  Briefly, as stellar activity creates magnetic hot spots in the chromosphere, emission of \ha and Na I photons is stimulated, filling in the observed absorption lines \citep[see][Figure 2]{robertson13}.

We define \ihap, our \ha index, according to \citet{robertson13}, using a method analogous to the calculation of \citet{kurster03}.  The index is simply a ratio of the flux in the 1.6 \AA~window centered on the \ha line relative to the nearby pseudocontinuum.  The index can easily be transformed into an equivalent width, and to \lhap, the ratio of the \ha luminosity to the stellar bolometric luminosity \citep[via][]{walkowicz04}.  Our Na I D index, \idp, is defined as described in \citet{diaz07}.  Like \ihap, \id is a ratio of the flux in the cores of the Na I D lines (1 \AA~windows centered on each line) to the adjacent pseudocontinuum.  We note that although \citet{gds11} report stronger correlation between their Na I D and Ca H \& K indices when using 0.5 \AA~windows for \idp, we have retained the 1 \AA~windows in order to compute \rpdp, the temperature-independent ratio of Na I D luminosity to bolometric luminosity \citep{diaz07}.  We note that when using the 0.5 \AA~windows we see essentially identical stellar magnetic behavior for GJ 328, but at lower signal-to-noise.  As a control test, we have also analyzed the Ca I ($\lambda = 6572.795$ \AA) line, which does not respond to stellar activity, and should therefore remain relatively constant.

\begin{figure}
\begin{center}
\subfigure[\label{act_all}]{\includegraphics[scale=0.45]{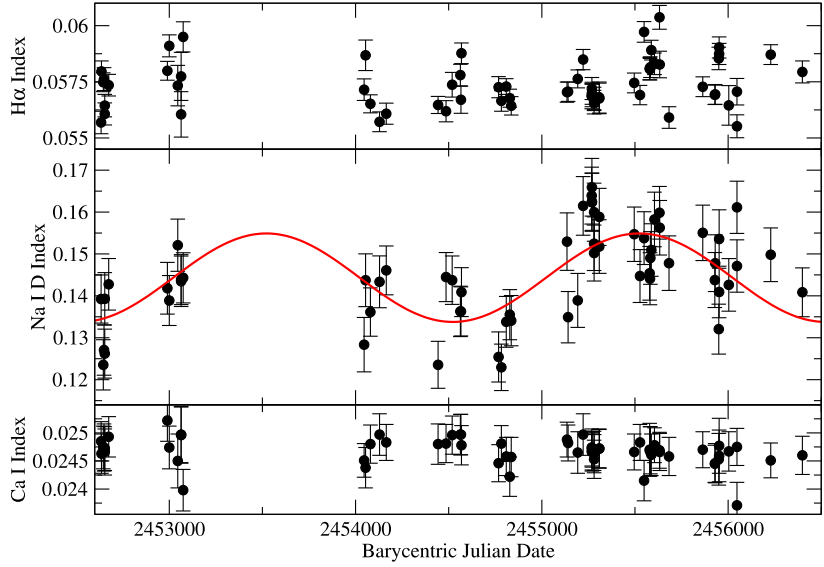}}
\subfigure[\label{act_phase}]{\includegraphics[scale=0.45]{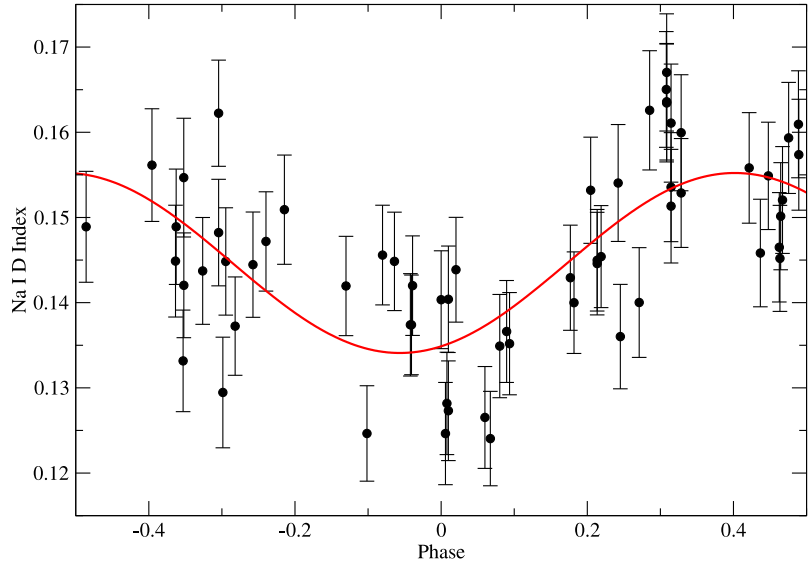}}
\caption{We monitor the stellar magnetic activity of GJ 328 through the variable depths of activity-sensitive absorption lines.  In (\emph{a}), we show our \ha (top) and Na I D (middle) indices at the time of each RV observation.  For the sodium index, we detect a periodic activity cycle with a period of 2000 days, shown in red.  We show the \id index folded to the period of the cycle in (\emph{b}).  The Ca I index (\emph{a}, bottom) is insensitive to stellar activity, and serves as a control measurement.}
\label{act}
\end{center}
\end{figure}

\begin{figure}
\begin{center}
\includegraphics[scale=0.5]{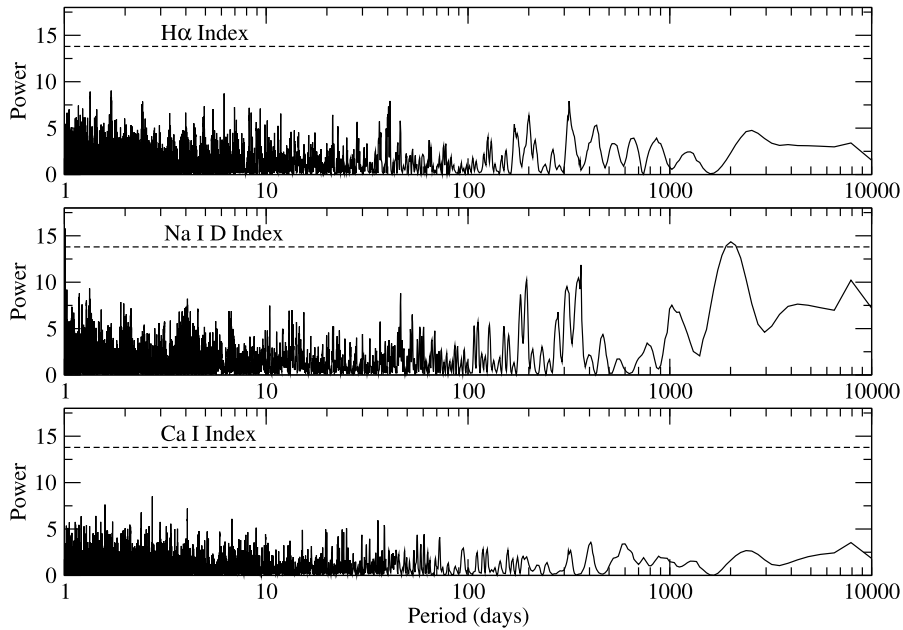}
\caption{Fully generalized Lomb-Scargle periodograms for the stellar activity indices shown in Figure \ref{act}.  The dashed horizontal lines represent the power level corresponding to a false alarm probability (FAP) of 0.01, as calculated from Equation 24 of \citet{zk09}.  We note that these FAP levels represent a preliminary estimate, and our formal FAP values are obtained through a bootstrap analysis, which we describe in Section \ref{analysis}.}
\label{act_ps}
\end{center}
\end{figure}

\begin{figure}
\begin{center}
\includegraphics[scale=0.5]{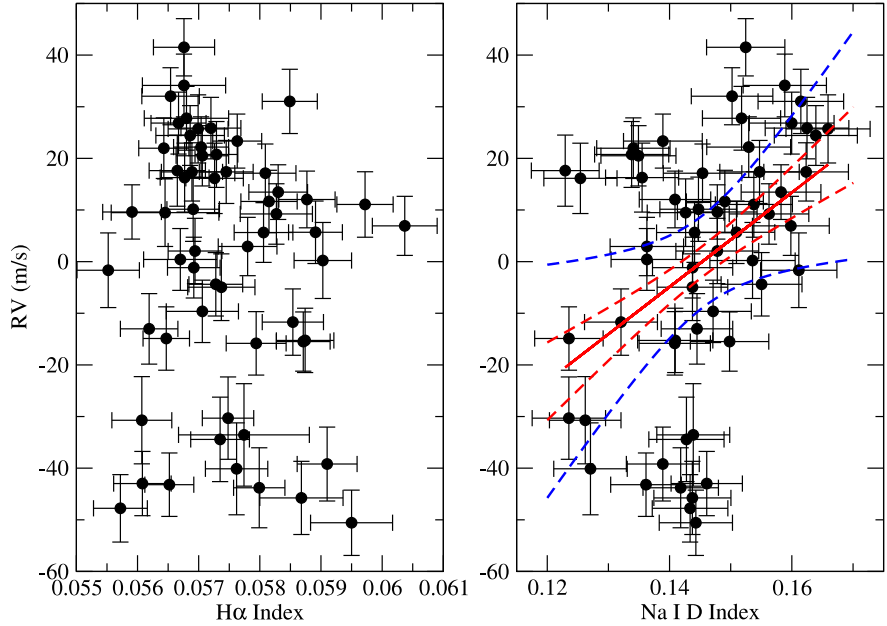}
\caption{To evaluate the influence of stellar activity on our RV measurements, we plot our HET/HRS RV measurements as a function of \iha (left) and \id (right).  We find that our RVs are correlated with \id at a statistically significant level.  Our best linear least-squares fit to the relation is shown as a solid red line.  The dashed curves indicate our $1\sigma$ (red) and $3\sigma$ (blue) error bounds on the fit.  Note that the planetary signal has not been removed from these data.}
\label{rv_act}
\end{center}
\end{figure}

GJ 328 has a mean \lha of -3.53, which is typical for a quiet star of comparable mass \citep{robertson13}.  We show our time series \iha data in the top panel of Figure \ref{act_all}.  To evaluate any periodic activity, we have computed a generalized Lomb-Scargle periodogram for \iha (Figure \ref{act_ps}, top).  We see no evidence in \ha for periodic behavior that might appear as a Keplerian signal in our RV data.  Furthermore, Figure \ref{rv_act} shows our measured RVs as a function of \ihap.  We find no correlation between the values, and therefore conclude there is no evidence in the \ha activity of GJ 328 that the observed planetary signal is caused by stellar activity.

Our analysis of the sodium D features requires more careful scrutiny in the context of interpreting the RV measurements.  We present our Na I D measurements in the middle panel of Figure \ref{act_all}.  The periodogram (middle panel, Figure \ref{act_ps}) shows a strong peak at 2006 days with a preliminary FAP of 0.006 \citep[as computed from][]{sturrock10}.  Our bootstrap false alarm routine produced no false alarms in $10^4$ trials, yielding an upper-limit final FAP of $10^{-4}$.  We consider this a highly significant detection for a stellar cycle, which may not be strictly periodic, and which may also be subject to stochastic activity in addition to cyclic behavior.  We fit a sinusoid of the form $I_{\textrm{D}}(t) = a_0 + a_1 \sin(\omega t + \phi)$ to the data, where $\omega = \frac{2\pi}{P}$, $P$ is the period of the cycle, and $a_0,~a_1$, and $\phi$ are free parameters to set the \id zero point, amplitude, and phase, respectively.  Our final fitted period is 2013 days, in good agreement with the periodogram peak, and has an amplitude of 0.0104 ($7.2\%$) in \idp.  The fit to the cycle is shown in Figure \ref{act}.

Given the presence of a stellar activity cycle for GJ 328, it is especially important to verify the observed exoplanet signature is not produced by this stellar magnetic behavior.  The right panel of Figure \ref{rv_act} shows our RVs as a function of \idp.  A Pearson correlation test yields a correlation coefficient of 0.41 which, for a sample size of 58 data points, indicates a probability $p = 0.0007$ of no correlation.  Evidently, then, our measured velocities include a component from the star itself.  Performing a standard linear least squares fit to the data, we find a relation

\begin{equation}
\label{id_rv}
\textrm{RV} (\textrm{m/s}) = -133 + 9.15 \times (I_{\textrm{D}}/0.01)
\end{equation}

\noindent with $1\sigma$ errors of 40 m/s on the intercept and 2.7 $\frac{\textrm{m/s}}{I_{D}/0.01}$ on the slope.

While we wish to verify the observed exoplanet signal is not caused by stellar activity, it is equally important to confirm that the \idp-RV correlation is not in fact created by the planet signal.  Since the periods of the planet and the cycle are near the 2:1 ratio, it could be the case we are simply seeing the first harmonic of the planet's period.  However, if the observed correlation were simply due to the high/low extremes of the exoplanet signal coincidentally occurring during periods of high/low stellar activity, we should expect to see a similar effect in \hap.  Since we do not, we conclude the correlation seen in Figure \ref{rv_act} is truly due to stellar activity shifting the stellar absorption lines.

%RV vs Na I (1A): Correlation coefficient = 0.396 => p=0.001.
%Fit: RV = -126.43 + 870.92*NaI
%Errors: Intercept=40, Slope=272
%Based on this, and based on the agreement bet. Residual RV &
%activity cycle, we conclude stellar RV amplitude must be
%between 8.4 and 16 m/s (Na amplitude * (slope +/- uncertainty))
%Be conservative, set to minimum.

%RV residual (to uncorrected RVs - fit from decorrelated RVs) vs Na (1A):
%Correlation coefficient = 0.66 => p=1E-8.
%Fit: RVres= -78.665 + 536.37*NaI
%Errors: Intercept=12.0, Slope=82.4

%GJ 184 CORRELATION:
%Fit: RV= -77.462 + 519.72*NaI
%Errors: Intercept=17.8, Slope=119
%Correlation coefficient=0.57 => p=4E-5 for 42 points.

\subsection{Stellar Activity Correction}

To investigate the effect of stellar activity on our velocity measurements, and its possible consequences for the existence of the planet, we subtracted the fit shown in Equation \ref{id_rv} from our HRS RVs.  We adopt a linear model because it fits within the physical interpretation that the RV-activity dependence is the result of convective redshift.  The process is described fully in \citet{kurster03}, who observe a similar dependence for Barnard's star (= GJ 699) on the \ha line.  Briefly, the outward convective motion of gas at the stellar photosphere typically produces a net blueshift of the measured absorption lines.  During periods of high stellar activity (increasing chromospheric emission and yielding higher \id values), regions of magnetic plage will suppress the local convection, resulting in a perceived redshift (i.e. a positive RV, hence the positive slope in Figure \ref{rv_act}).  We therefore expect the convective redshift to increase roughly linearly as the stellar magnetic activity increases.

\begin{figure}
\begin{center}
\includegraphics[scale=0.5]{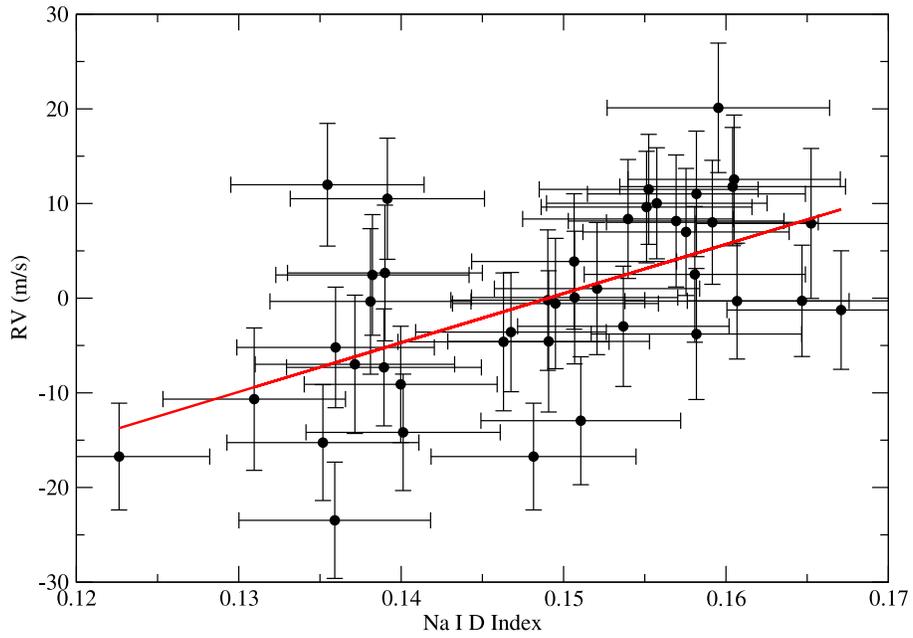}
\caption{RV as a function of \id for GJ 184, another M0 star in our M dwarf survey.  As for GJ 328 and GJ 699, RV increases at higher stellar activity.  We interpret this phenomenon as resulting from magnetic plage suppressing local photospheric convection during periods of higher activity.  Our best linear least-squares fit to the data is shown as a red line.}
\label{gj184}
\end{center}
\end{figure}

To illustrate the effect of convective redshift in a simpler case, we show in Figure \ref{gj184} our measured RVs for GJ 184 versus \idp.  GJ 184, another M0 dwarf in our survey, has no known exoplanets, thereby eliminating the large scatter created by the presence of a giant planet in the velocity data.  We see again that RV increases as a function of sodium emission, and the relation is nicely approximated as linear.  In the case of GJ 184, we find a linear dependence of RV(m/s) = $-77 + 5.20 \times (I_{\textrm{D}}/0.01)$, the slope of similar order to that found for GJ 328.

\begin{figure}
\begin{center}
\includegraphics[scale=0.5]{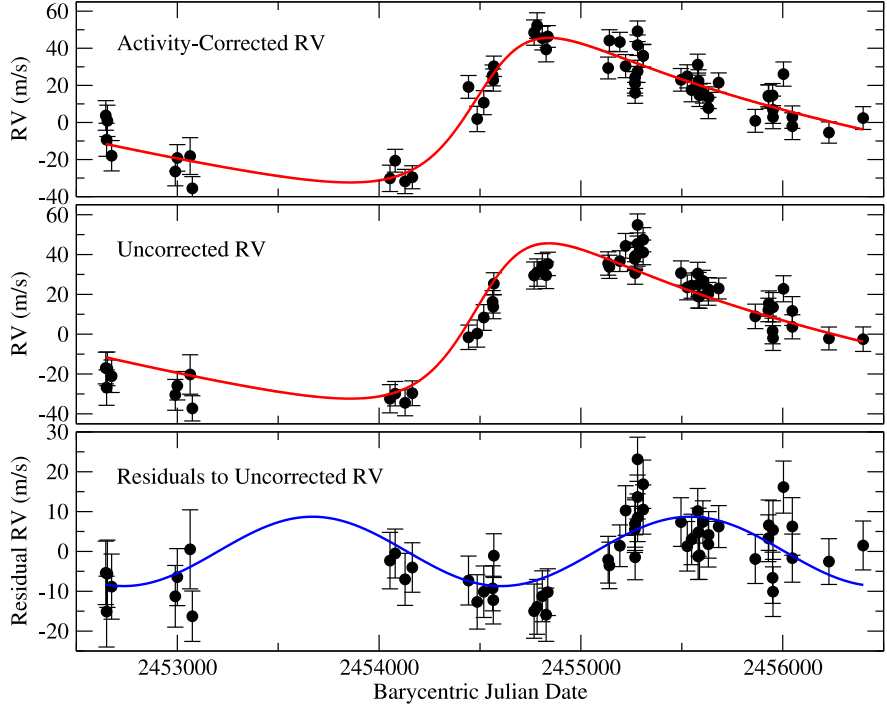}
\caption{Upon subtracting the relation in Equation \ref{id_rv} from our HET/HRS RVs, we obtain the corrected velocities shown in the top panel.  We perform a Keplerian fit to these RVs, shown in red.  When subtracting this fit from the uncorrected RVs (middle panel), we recover an 1865-day signal in the residual velocities (bottom).  The period and phase of this signal (fit shown in blue) are consistent with the activity cycle observed for the Na I D features.}
\label{act_fit}
\end{center}
\end{figure}

The ``corrected" RVs for GJ 328 are shown in Figure \ref{act_fit}; from inspection, it is clear that a large signal is still present after removing the activity correlation.  A periodogram of the de-trended data again shows a strong peak at 3500 days, with a bootstrap FAP less than $10^{-4}$.  Performing a Keplerian fit to the corrected velocities yields orbital parameters largely consistent with those of our original fit, although the eccentricity increases from its original value of 0.29 to 0.44.

We wish to verify that the observed correlation between RV and \id is a result of the stellar activity cycle identified herein.  To do so, we subtracted our orbital fit to the activity-corrected RVs from the uncorrected velocities (Figure \ref{act_fit}, middle panel).  A periodogram of the residuals reveals another strong peak at 1830 days, with a bootstrap FAP again falling below $10^{-4}$.  Fitting a circular orbit to the residuals (so as to preserve the assumption of a sinusoidal fit to the activity cycle), we find a period of 1870 days with mean anomaly $M_0 = 163^{\circ}$.  The fit has an RV amplitude of 8.7 m/s.  We show the residual velocities and the associated fit in the bottom panel of Figure \ref{act_fit}.  Returning to our time-series \id data, we performed a second sinusiodal fit, with the period and phase fixed to match the residual RV signal.  This fit yields an RMS scatter of $0.00822$ in \id, compared to $0.00796$ for the 2013-day fit.  Performing an F-test to compare the fits, we find a probability $P = 0.79$ that the fits are equally valid.  Therefore, our observed 1870-day RV signal is statistically consistent with the activity cycle present in the sodium D features.  Furthermore, upon applying the planetary fit obtained from the ``corrected" velocities (shown in Figure \ref{act_fit}) to the uncorrected RVs, the Pearson correlation coefficient between the resulting residual RVs and \id increases to 0.66, corresponding to a probability $p < 10^{-8}$ of no correlation.  Based on these tests, we conclude that the correlation between RV and \id is the result of apparent stellar velocity shifts caused by the periodic activity cycle of GJ 328.

Properly correcting for the influence of stellar activity on our RVs is problematic.  Due to the presence of planet b's signal, there is a high amount of scatter in the RV-versus-\id relation, which leads to large uncertainties in Equation \ref{id_rv}.  It is therefore not wise to assume that the ``corrected" velocities shown in Figure \ref{act_fit} are properly adjusted.  It is also not possible to perform a two-signal fit with the period and phase of the activity cycle fixed.  Given our current data set, the eccentric signal of planet b and the sinusoidal activity cycle can be easily modeled as a single, mildly eccentric Keplerian.  Therefore, any two-signal fit where the amplitude of the activity cycle is allowed to vary will result in an amplitude of zero for the cycle.  While such a model is preferable statistically, it does not account for the additional information contained in the \id data.

\begin{figure}
\begin{center}
\includegraphics[scale=0.5]{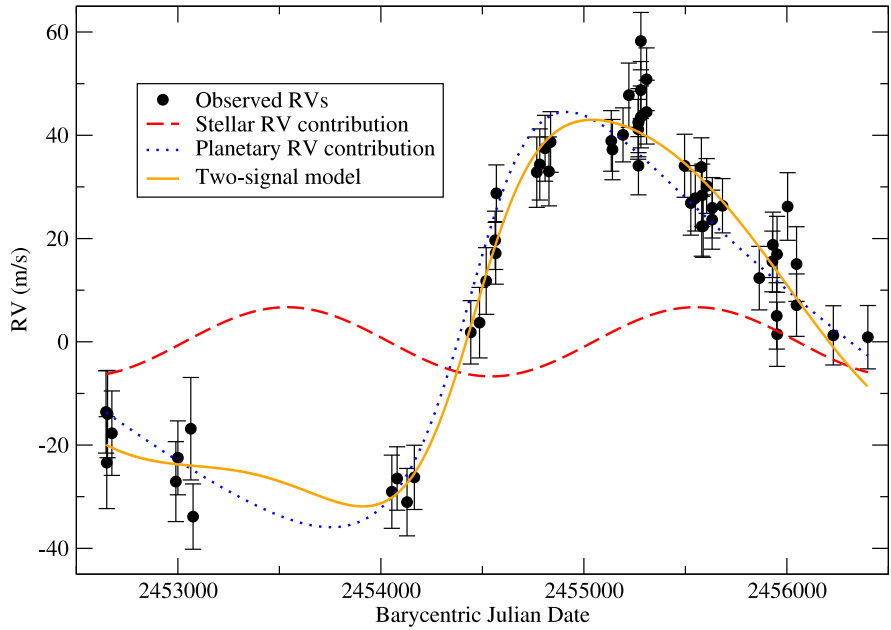}
\caption{HET/HRS RVs for GJ 328, showing our two-signal model derived from the RV and \id data.  The solid orange curve indicates the model to the data, while the dashed red and dotted blue curves show the individual RV contributions from the stellar magnetic cycle and planet b, respectively.}
\label{2fit}
\end{center}
\end{figure}

Because of the difficulties listed above, we have elected to take a conservative approach in modeling the system based on our current data.  From the slope of Equation \ref{id_rv} and the amplitude of the \id cycle, we expect the RV amplitude of the stellar cycle to be between 6.7 and 11.9 m/s.  We have therefore fixed the RV amplitude for the stellar cycle to 6.7 m/s, which we consider the minimum possible given our analysis.  Also, rather than retain the 1870-day period for the cycle, we have fixed its period and phase to the 2013-day solution found for the \id data, since we expect that data to be free of any influence from planet b's signal.  Using these assumptions for the stellar cycle, we subtracted the stellar activity signal from our HRS RVs and re-modeled the orbit of planet b.  Our final ``corrected" model is given alongside the uncorrected model in Table \ref{orbit}, and we show the model, decomposed into the stellar and planetary components, in Figure \ref{2fit}.  

We note that for this conservative treatment of the stellar activity, the orbital parameters typically only differ by approximately $1\sigma$.  It is perhaps more interesting to examine \emph{how} the orbit of planet b changes as the activity cycle's RV amplitude increases, rather than to what magnitude.  Most notably, the planet's eccentricity continues to increase as we assign higher amplitudes to the stellar RV contribution.  In general, we can say confidently that because of the activity-induced component to our measured RVs, planet b is more eccentric--and consequently less massive--than implied by a simple single-Keplerian fit to the uncorrected velocities.

\section{\bf Discussion}

GJ 328b joins the rapidly growing list of long-period giant planets emerging as the McDonald Observatory exoplanet survey approaches a decade of semi-constant monitoring on many of its targets \citep{robertson12a,robertson12b}.  Such discoveries illustrate the importance of the long-term RV surveys in obtaining a complete census of the Galactic planet population.  Transit surveys such as \emph{Kepler} \citep{borucki10} will not operate long enough to find planets in Jupiter-like orbits, and imaging programs are currently unable to observe planets at $a \lesssim 10$ AU.  While microlensing \citep[e.g.][]{gould10} is sensitive to Jupiter analogs, the possibility of detailed characterization for both the star and planet are extremely limited.

The large mass and orbital separation of GJ 328b offer the potential of further study via astrometry or imaging.  Adopting our activity-corrected orbital fit, we calculate an amplitude $\alpha \sin i = 0.70$ mas for the astrometric motion of GJ 328.  Such motion is well within the detection limits of the Fine Guidance Sensor on HST \citep{nelan10}.  However, the long orbital period likely makes an astrometric campaign prohibitively expensive.  Similarly, the sky-projected separation of the planet is approximately 220 mas, slightly more than half the 368-mas separation of HR 8799e \citep{marois10}, which might be resolvable for an M star.  Unlike HR 8799, though, the lack of X-ray emission \citep{hunsch99} or rotational line broadening indicates GJ 328 is an old star, and the planet will therefore be cold.  The resulting lack of thermal emission from the planet should render direct imaging impossible for current instruments.

On the other hand, whereas all of our previously published long-period giant planets have been found around solar-type stars, GJ 328b is unique in that it orbits a red dwarf star.  It is currently the most massive and most distant planet found to orbit a low-mass star\footnote{Excluding planets found via gravitational microlensing, due to the large uncertainties in stellar and orbital parameters.}.  Along with GJ 832b \citep{bailey09}, it is one of only two M dwarf planets with $P \gtrsim 10$ years.  While theoretical analyses predicting a deficit of giant planets at small orbital radii around M stars have been thoroughly confirmed by observation \citep{endl06}, further study is required to determine whether GJ 832 and GJ 328 are anomalies, or whether the Jovian population of low-mass stars is more similar to their FGK counterparts at larger separation.  Interestingly, both stars fail to show super-Solar metal content; using the \citet{schlaufman10} calibration, GJ 832 has [M/H] = -0.24, while GJ 328 is roughly solar at [M/H] = 0.00.  These metallicities may be considered low in light of the well-established metallicity-frequency relation for giant planets \citep{fischer05}, which is generally seen as strong evidence of planet formation via core accretion.  Considering both planets fall within the range of semi-major axes where \citet{boss06} shows gas giants can quickly form via gravitational instability around M dwarfs, the lack of metal excess could be seen as evidence that these planets formed via direct gravitational collapse.  Such a claim would be strengthened with the discovery of additional Jupiter analogs around ``metal-poor" M dwarfs.

Regardless of the formation mechanism, it appears that low-mass stars can form giant planets even without a highly abundant supply of heavy elements.  This seems to rule out the possibility that our survey found no close-in gas giants around M dwarfs because our targets are biased towards metal-poor stars \citep{endl06}.  The relatively low frequency of Jovian planets inside 1 AU around low-mass stars must therefore either reflect an overall underabundance of large planets relative to FGK stars, or point towards a mechanism preventing inward planetary migration for cool stars.

The presence of a long-period ($P \sim 5$ years) solar-type cycle in GJ 328 adds further evidence that activity cycles are commonplace amongst red dwarf stars \citep[e.g.][]{buccino11,gds12,robertson13}.  GJ 328's relatively high mass for an M star is consistent with the current understanding that stars massive enough to maintain a radiative inner envelope--and thus a tachocline--are likely to exhibit activity cycles, while fully convective stars will not \citep[see][and references therein]{robertson13}.  This trend seems to point to the ubiquity of the tachocline-driven magnetic dynamo for maintaining solar-like magnetic activity in old main-sequence stars.  As the period of the activity cycle is too long to be the result of spot modulation via stellar rotation, we conclude we are observing cycles in the mean granulation pattern on the stellar surface.  The resulting effect on RV must therefore be due to variations in the percentage of the chromosphere covered by cells of hot gas, convecting upward and creating a net blueshift.

The appearance of GJ 328's activity cycle in the Na I D resonance lines and not in \ha reaffirms the conclusions of \citet{diaz07} and \citet{gds11} that \id is the most sensitive tracer of stellar activity in low-mass stars for spectrographs which do not acquire Ca II H\&K.  We are in the process of investigating \id variability for our entire M dwarf data set, and will soon have a more quantitative comparison between the \id and \iha tracers.

While it is not common for stellar magnetic activity to create an RV signal with $K \gtrsim 8$ m/s, we have previously identified two such stars within our M dwarf sample \citep{robertson13}, suggesting such behavior is not highly unusual.  Unfortunately, our time sampling for GJ 328 prevents us from obtaining a fully quantitative two-signal model for our RVs.  However, our current data set still provides some insight as to what effect stellar activity has on our derived properties for planet b.  The positive correlation between RV and \id (Figure \ref{rv_act}) ensures that RV should change in phase with \idp, rather than the two quantities being a half cycle out of phase.  As a result, the ``true" orbit of planet b will always be more eccentric than implied by a single-Keplerian fit to our data for any amplitude of the activity cycle.  Because more eccentric orbits have higher RV amplitudes at fixed $a$, the planet must also be less massive than found in our uncorrected fit.  Still, the difficulty inherent in separating stellar and planetary RV signals even for a planet with $K > 40$ m/s illustrates the need to exercise a great deal of caution when considering planets with RV amplitudes comparable to (or smaller than) signals caused by stellar activity \citep[e.g.][]{dumusque12}.  We see also that period commensurability between stellar and planetary signals need not automatically disqualify an RV signal as an exoplanet, as the periods of the planet and cycle for GJ 328 are near 2:1.  In the case of the Sun (activity cycle period $\simeq 11$ years) and Jupiter ($P = 12$ years), the periods of planets and activity cycles may be very close to 1:1 commensurability.  A planetary signal need not be disregarded because RV measurements correlate with activity indices, or because the stellar activity displays periodic behavior.  In such cases, though, it is doubly important to include a thorough analysis of stellar magnetic behavior before accepting any planetary solution.

\section{\bf Conclusions}

We have discovered a ``cold Jupiter" planet orbiting approximately 4.5 AU from the late K/early M star GJ 328.  Like many old dwarfs, GJ 328 exhibits a long-period magnetic cycle, which we see in the variability of the Na I D lines.  We have shown that this activity cycle influences our measured RVs.  Although we are unable to make a statistically robust two-signal model that accounts for both the stellar and planetary velocity contributions, we show that the fit to planet b must become more eccentric as the RV amplitude of the stellar cycle increases.

\begin{acknowledgements}

P. R. is supported by a University of Texas Graduate Continuing Fellowship.  M. E. and W. D. C. acknowledge support from the National Aeronautics and Space Administration through the Origins of Solar Systems Program.  This work also benefits from previous NASA grants NNX07AL70G and NNX10AL60G.  The Hobby-Eberly Telescope (HET) is a joint project of the University of Texas at Austin, the Pennsylvania State University, Stanford University, Ludwig-Maximilians-Universit\"{a}t M\"{u}nchen, and Georg-August-Universit\"{a}t G\"{o}ttingen.  The HET is named in honor of its principal benefactors, William P. Hobby and Robert E. Eberly.  We would like to thank the McDonald Observatory TAC for generous allocation of observing time.  We are grateful to the HET Resident Astronomers and Telescope Operators for their valuable assistance in gathering our HET/HRS data.  This research has made use of the Exoplanet Orbit Database and the Exoplanet Data Explorer at exoplanets.org.

\end{acknowledgements}

\clearpage

\begin{table}
\begin{center}

\begin{tabular}{| l l | l |}
\hline & & \\
Parameter & Value & Reference \\
\hline & & \\
Spectral Type & M1 d & \citet{kharchenko09} \\
$V$ & $9.98 \pm 0.04$ & \citet{hog00} \\
$B-V$ & $1.32 \pm 0.1$ & \citet{hog00} \\
$K$ & $6.352 \pm 0.026$ & \citet{cutri03} \\
$M_{V}$ & $8.50 \pm 0.13$ & This Work \\
$M_{K}$ & $4.87 \pm 0.06$ & This Work \\
Parallax & $50.52 \pm 1.90$ mas & \citet{vanl07} \\
Distance & $19.8 \pm 0.8$ pc & This Work \\
T$_{eff}$ & $3900 \pm 100$ K & \citet{morales08} \\
%$\log g$ & $4.45 \pm 0.12$ \\
$[$M/H$]$ & $0.00 \pm 0.15$ & \citet{schlaufman10} \\
%$\xi$ & $1.20 \pm 0.15$ km/s \\
Mass & $0.69 \pm 0.05 M_{\odot}$ & \citet{delfosse00} \\
$[L_{\textrm{H}\alpha}/L_{\textrm{bol}}]$ & $-3.53 \pm 0.01$ & \citet{walkowicz04} \\
$R'_{\textrm{D}}$ & $-4.91 \pm 0.05$ & \citet{diaz07} \\
%Age\tablenotemark3 & 7.20 Gyr \\
%$\log R'_{HK}$ & $-4.65 \pm 0.03$ \\
\hline
\end{tabular}
\caption{Stellar Properties for GJ 328}
\label{stellar}

\end{center}
\end{table}

\begin{table}
\begin{center}

\begin{tabular}{| l l l |}
\hline & & \\
Orbital Parameter & Value & Value \\
 & (Uncorrected for Stellar Activity) & (Corrected for Stellar Activity) \\
\hline & & \\
Period $P$ (days) & $4100 \pm 170$ & $4100 \pm 300$ \\
Periastron Passage $T_0$ & $4600 \pm 70$ & $4500 \pm 100$ \\
(BJD - 2 450 000) & & \\
RV Amplitude $K$ (m/s) & $42 \pm 1.7$ & $40 \pm 2.0$ \\
Mean Anomaly $M_0$\tablenotemark3 & $190^{\circ} \pm 9^{\circ}.0$ & $200^{\circ} \pm 9^{\circ}.0$ \\
Eccentricity $e$ & $0.29 \pm 0.04$ & $0.37 \pm 0.05$ \\
Longitude of Periastron $\omega$ & $290^{\circ} \pm 8^{\circ}.0$ & $290^{\circ} \pm 3^{\circ}.0$ \\
Semimajor Axis $a$ (AU) & $4.4 \pm 0.30$ & $4.5 \pm 0.20$ \\
Minimum Mass $M \sin i$ ($M_{J}$) & $2.5 \pm 0.10$ & $2.3 \pm 0.13$ \\
HET/HRS RV offset (m/s) & $-18.0$ & $-17.0$ \\
HJST/Tull RV offset (m/s) & $-18.0$ & N/A \\
Keck/HIRES RV offset (m/s) & $-7.2$ & N/A \\
RMS (m/s) & 6.0 & 6.0 \\
%Stellar ``jitter'' (m/s) & & \\
\hline
\end{tabular}
\caption{Orbital solutions for GJ 328b, with and without corrections for stellar activity.}
\label{orbit}
\tablenotetext{3}{Evaluated at the time of the first RV point in Table \ref{rv}}
\end{center}
\end{table}

\clearpage

\begin{center}

\footnotesize

\tablecaption{Radial Velocities and Stellar Activity Indices for GJ 328}
\label{rv}	
\tablefirsthead{\hline
BJD - 2450000 & Radial Velocity & Uncertainty & \iha & \id \\
& (m/s) & (m/s) & & \\
\hline}
		
\tablehead{\hline
\emph{Table \ref{rv} cont'd.} & & & & \\ \hline
BJD - 2450000 & Radial Velocity & Uncertainty & \iha & \id \\
& (m/s) & (m/s) & & \\
\hline}
		
\tabletail{\hline}

\begin{supertabular}{| l l l l l |}

\multicolumn{5}{| l |}{\emph{HET/HRS Velocities}} \\
2645.81037607 & 	 -30.31 & 	 7.97 & 	 $ 0.05748 \pm 0.00042 $ & 	 $ 0.12353 \pm 0.00600 $ \\
2649.79770587 & 	 -40.13 & 	 8.90 & 	 $ 0.05762 \pm 0.00051 $ & 	 $ 0.12707 \pm 0.00601 $ \\
2653.79836840 & 	 -30.72 & 	 8.45 & 	 $ 0.05644 \pm 0.00050 $ & 	 $ 0.13929 \pm 0.00625 $ \\
2674.73658463 & 	 -34.43 & 	 8.18 & 	 $ 0.05735 \pm 0.00048 $ & 	 $ 0.14276 \pm 0.00615 $ \\
2989.98128744 & 	 -43.80 & 	 7.74 & 	 $ 0.05799 \pm 0.00042 $ & 	 $ 0.14182 \pm 0.00617 $ \\
2999.96108900 & 	 -39.20 & 	 7.16 & 	 $ 0.05910 \pm 0.00049 $ & 	 $ 0.13889 \pm 0.00596 $ \\
3063.78296966 & 	 -33.56 & 	 9.92 & 	 $ 0.05774 \pm 0.00107 $ & 	 $ 0.14387 \pm 0.00596 $ \\
3074.76013011 & 	 -50.58 & 	 6.33 & 	 $ 0.05950 \pm 0.00067 $ & 	 $ 0.14430 \pm 0.00599 $ \\
4053.95303460 & 	 -45.76 & 	 7.09 & 	 $ 0.05868 \pm 0.00068 $ & 	 $ 0.14373 \pm 0.00630 $ \\
4079.87702463 & 	 -43.20 & 	 6.14 & 	 $ 0.05652 \pm 0.00040 $ & 	 $ 0.13614 \pm 0.00577 $ \\
4128.87683185 & 	 -47.78 & 	 6.53 & 	 $ 0.05572 \pm 0.00044 $ & 	 $ 0.14335 \pm 0.00618 $ \\
4164.77031060 & 	 -42.98 & 	 6.22 & 	 $ 0.05608 \pm 0.00047 $ & 	 $ 0.14608 \pm 0.00583 $ \\
4442.99734598 & 	 -14.89 & 	 6.14 & 	 $ 0.05647 \pm 0.00038 $ & 	 $ 0.12354 \pm 0.00560 $ \\
4485.88562946 & 	 -13.02 & 	 6.83 & 	 $ 0.05619 \pm 0.00047 $ & 	 $ 0.14448 \pm 0.00585 $ \\
4518.68369633 & 	 -4.95 & 	 6.46 & 	 $ 0.05737 \pm 0.00055 $ & 	 $ 0.14374 \pm 0.00578 $ \\
4562.66614754 & 	 2.93 & 	 5.65 & 	 $ 0.05780 \pm 0.00048 $ & 	 $ 0.13629 \pm 0.00601 $ \\
4565.65834052 & 	 0.42 & 	 5.99 & 	 $ 0.05670 \pm 0.00060 $ & 	 $ 0.13629 \pm 0.00585 $ \\
4568.66504682 & 	 12.04 & 	 5.49 & 	 $ 0.05877 \pm 0.00046 $ & 	 $ 0.14089 \pm 0.00584 $ \\
4767.99308233 & 	 16.13 & 	 6.81 & 	 $ 0.05726 \pm 0.00047 $ & 	 $ 0.12542 \pm 0.00598 $ \\
4782.95101189 & 	 17.62 & 	 6.89 & 	 $ 0.05665 \pm 0.00046 $ & 	 $ 0.12294 \pm 0.00554 $ \\
4808.89548004 & 	 20.75 & 	 6.38 & 	 $ 0.05729 \pm 0.00035 $ & 	 $ 0.13380 \pm 0.00606 $ \\
4827.84051511 & 	 16.28 & 	 6.68 & 	 $ 0.05677 \pm 0.00040 $ & 	 $ 0.13551 \pm 0.00598 $ \\
4835.92477666 & 	 21.97 & 	 5.87 & 	 $ 0.05643 \pm 0.00041 $ & 	 $ 0.13408 \pm 0.00600 $ \\
5134.99081805 & 	 22.18 & 	 5.87 & 	 $ 0.05704 \pm 0.00045 $ & 	 $ 0.15294 \pm 0.00686 $ \\
5140.96822648 & 	 20.50 & 	 5.84 & 	 $ 0.05706 \pm 0.00043 $ & 	 $ 0.13491 \pm 0.00613 $ \\
5192.95380323 & 	 23.35 & 	 5.23 & 	 $ 0.05763 \pm 0.00040 $ & 	 $ 0.13891 \pm 0.00644 $ \\
5221.76209389 & 	 31.03 & 	 6.22 & 	 $ 0.05849 \pm 0.00045 $ & 	 $ 0.16147 \pm 0.00700 $ \\
5267.63490551 & 	 24.43 & 	 5.79 & 	 $ 0.05686 \pm 0.00056 $ & 	 $ 0.16392 \pm 0.00679 $ \\
5268.75910493 & 	 25.86 & 	 5.95 & 	 $ 0.05720 \pm 0.00049 $ & 	 $ 0.16248 \pm 0.00684 $ \\
5268.76694488 & 	 17.38 & 	 5.64 & 	 $ 0.05689 \pm 0.00050 $ & 	 $ 0.16233 \pm 0.00692 $ \\
5268.77478240 & 	 25.70 & 	 6.60 & 	 $ 0.05699 \pm 0.00049 $ & 	 $ 0.16592 \pm 0.00688 $ \\
5280.70816761 & 	 41.51 & 	 5.55 & 	 $ 0.05676 \pm 0.00050 $ & 	 $ 0.15245 \pm 0.00640 $ \\
5280.71600501 & 	 32.03 & 	 5.52 & 	 $ 0.05654 \pm 0.00047 $ & 	 $ 0.15022 \pm 0.00667 $ \\
5280.72383987 & 	 26.83 & 	 5.99 & 	 $ 0.05667 \pm 0.00045 $ & 	 $ 0.15997 \pm 0.00693 $ \\
5308.63116619 & 	 34.12 & 	 6.07 & 	 $ 0.05676 \pm 0.00068 $ & 	 $ 0.15884 \pm 0.00681 $ \\
5308.63906025 & 	 27.76 & 	 6.18 & 	 $ 0.05680 \pm 0.00069 $ & 	 $ 0.15177 \pm 0.00639 $ \\
5496.00271940 & 	 17.37 & 	 6.11 & 	 $ 0.05745 \pm 0.00044 $ & 	 $ 0.15471 \pm 0.00649 $ \\
5526.92170976 & 	 10.15 & 	 6.22 & 	 $ 0.05691 \pm 0.00043 $ & 	 $ 0.14472 \pm 0.00631 $ \\
5548.87054184 & 	 11.07 & 	 6.34 & 	 $ 0.05972 \pm 0.00046 $ & 	 $ 0.15378 \pm 0.00630 $ \\
5578.88402982 & 	 17.12 & 	 5.66 & 	 $ 0.05809 \pm 0.00050 $ & 	 $ 0.14539 \pm 0.00642 $ \\
5580.88847924 & 	 5.64 & 	 5.78 & 	 $ 0.05806 \pm 0.00048 $ & 	 $ 0.14409 \pm 0.00622 $ \\
5582.89142512 & 	 11.65 & 	 6.04 & 	 $ 0.05815 \pm 0.00046 $ & 	 $ 0.14904 \pm 0.00629 $ \\
5587.75750742 & 	 5.68 & 	 6.05 & 	 $ 0.05891 \pm 0.00044 $ & 	 $ 0.15094 \pm 0.00627 $ \\
5604.70581615 & 	 13.46 & 	 5.30 & 	 $ 0.05830 \pm 0.00048 $ & 	 $ 0.15823 \pm 0.00652 $ \\
5631.63868921 & 	 6.93 & 	 5.72 & 	 $ 0.06037 \pm 0.00053 $ & 	 $ 0.15983 \pm 0.00628 $ \\
5632.74633132 & 	 9.23 & 	 5.85 & 	 $ 0.05827 \pm 0.00059 $ & 	 $ 0.15626 \pm 0.00651 $ \\
5682.61355140 & 	 9.62 & 	 5.26 & 	 $ 0.05591 \pm 0.00048 $ & 	 $ 0.14780 \pm 0.00651 $ \\
5864.00738743 & 	 -4.39 & 	 6.15 & 	 $ 0.05728 \pm 0.00044 $ & 	 $ 0.15504 \pm 0.00661 $ \\
5928.82773310 & 	 -1.16 & 	 5.88 & 	 $ 0.05692 \pm 0.00043 $ & 	 $ 0.14377 \pm 0.00656 $ \\
5930.81296731 & 	 2.06 & 	 6.33 & 	 $ 0.05694 \pm 0.00044 $ & 	 $ 0.14780 \pm 0.00677 $ \\
5949.88927832 & 	 -11.71 & 	 6.43 & 	 $ 0.05854 \pm 0.00050 $ & 	 $ 0.13206 \pm 0.00596 $ \\
5951.75489283 & 	 0.22 & 	 7.36 & 	 $ 0.05903 \pm 0.00047 $ & 	 $ 0.15357 \pm 0.00697 $ \\
5951.87637749 & 	 -15.26 & 	 6.23 & 	 $ 0.05874 \pm 0.00047 $ & 	 $ 0.14093 \pm 0.00615 $ \\
6003.73677125 & 	 9.48 & 	 6.54 & 	 $ 0.05645 \pm 0.00088 $ & 	 $ 0.14262 \pm 0.00627 $ \\
6047.61554908 & 	 -1.67 & 	 7.23 & 	 $ 0.05552 \pm 0.00051 $ & 	 $ 0.16113 \pm 0.00623 $ \\
6047.62618002 & 	 -9.63 & 	 6.04 & 	 $ 0.05706 \pm 0.00059 $ & 	 $ 0.14712 \pm 0.00625 $ \\
6228.99908594 & 	 -15.48 & 	 5.74 & 	 $ 0.05871 \pm 0.00044 $ & 	 $ 0.14981 \pm 0.00641 $ \\
6398.63806982 & 	 -15.83 & 	 6.14 & 	 $ 0.05794 \pm 0.00049 $ & 	 $ 0.14085 \pm 0.00583 $ \\
\multicolumn{5}{| l |}{\emph{Keck/HIRES Velocities}} \\
5222.10054500 &          21.80 &         3.31 &          $ \cdots $ &    $ \cdots $ \\
5609.85679500 &          2.59 &          3.66 &          $ \cdots $ &    $ \cdots $ \\
5611.02903000 &          0.19 &          4.69 &          $ \cdots $ &    $ \cdots $ \\
6315.04761800 &          -24.58 &        2.94 &          $ \cdots $ &    $ \cdots $ \\
\multicolumn{5}{| l |}{\emph{HJST/Tull Velocities}} \\
5286.67118600 &          6.90 &          5.03 &          $ \cdots $ &    $ \cdots $ \\
5290.69557300 &          9.18 &          7.41 &          $ \cdots $ &    $ \cdots $ \\
5291.68657100 &          -2.69 &         4.89 &          $ \cdots $ &    $ \cdots $ \\
5341.63611800 &          6.55 &          4.67 &          $ \cdots $ &    $ \cdots $ \\
5347.63440000 &          12.63 &         6.12 &          $ \cdots $ &    $ \cdots $ \\
5469.98876500 &          0.72 &          6.12 &          $ \cdots $ &    $ \cdots $ \\
5493.97635400 &          15.75 &         5.18 &          $ \cdots $ &    $ \cdots $ \\
5496.98315600 &          -10.00 &        5.38 &          $ \cdots $ &    $ \cdots $ \\
5523.96230100 &          -4.21 &         5.47 &          $ \cdots $ &    $ \cdots $ \\
5528.96290100 &          4.57 &          7.77 &          $ \cdots $ &    $ \cdots $ \\
5529.94187300 &          -8.76 &         5.89 &          $ \cdots $ &    $ \cdots $ \\
5548.90295700 &          -5.46 &         3.61 &          $ \cdots $ &    $ \cdots $ \\
5615.71883100 &          -6.75 &         5.02 &          $ \cdots $ &    $ \cdots $ \\
5632.69279200 &          -18.46 &        5.95 &          $ \cdots $ &    $ \cdots $ \\

\end{supertabular}

\end{center}

\end{document}